\documentclass{jfm}
\usepackage{multirow}
\usepackage{graphicx}
\usepackage{natbib}
\usepackage{amsmath}
\usepackage{color,xcolor}
\usepackage{mathrsfs}
\DeclareMathAlphabet{\mathscrbf}{OMS}{mdugm}{b}{n}
\usepackage{amssymb}
\usepackage{dcolumn}
\usepackage{bm}
\usepackage{booktabs}
\usepackage{overpic}
\usepackage{hyperref}
\hypersetup{
    breaklinks=true,
    colorlinks=true,
    citecolor=blue,
    linkcolor=blue,
    pdfauthor=Youcheng Xi,
    pdfkeywords=Flow instability,
    pdfsubject=attachment-line,
    pdftitle=Receptivity/stability of hypersonic leading-edge sweep flows around a blunt body
    }
\usepackage[colorinlistoftodos,prependcaption,textsize=tiny,shadow]{todonotes}

\graphicspath{{Figures/}}

\shorttitle{Receptivity/stability of hypersonic leading-edge sweep flows around a blunt body}
\shortauthor{Y. C. Xi, J. Ren, L. Wang and S. Fu}

\title{Receptivity and stability of hypersonic leading-edge sweep flows around a blunt body}

\author{Youcheng Xi\aff{1}, Jie Ren\aff{2}, Liang Wang\aff{1}
 \and Song Fu\aff{1}\corresp{\email{fs-dem@tsinghua.edu.cn}}}

\affiliation{\aff{1}School of Aerospace Engineering, Tsinghua University, Beijing 100084, China
\aff{2}Department of Mechanical Engineering, Faculty of Engineering, University of Nottingham, Nottingham, NG7 2RD, UK}
\begin{document}

\maketitle
\begin{abstract}
This study performs global stability/receptivity analyses of hypersonic sweep flows around a blunt body with an infinite span. For the first time, we obtain the characteristics of the leading attachment-line mode to the variation of sweep angles from 20 to 70 degrees. The global eigenfunctions exhibit the characteristics of the attachment-line instability at the leading edge. At the same time, cross-flow (at small sweep angles) or second Mack mode (at larger sweep angles) dominates further downstream.  We establish an adjoint-based bi-orthogonal eigenfunction system to address the receptivity problem of such flows to any external forces and boundary perturbations. The receptivity analyses indicate that the global modes are the most responsive to external forces and surface perturbations applied in the vicinity of the attachment line, regardless of the sweep angles. It is also proved that the present global extension of the bi-orthogonal eigenfunction system can be successfully applied to complex flows.
\end{abstract}


\section{Introduction} \label{S1}
The studies of three-dimensional sweep boundary layers date back to years ago. Most of them are based on local models: the attachment-line models and three-dimensional cross-flow models.
The sweep flow at leading-edge is often modeled by the sweep Hiemenz configuration, and the most unstable mode is symmetric along the chordwise direction perpendicular to the attachment line \citep{Lin1996,Theofilis1998,Obrist2003b}. Further downstream, because of the non-alignment in three-dimensional inviscid streamlines and pressure gradients, an inflection point appears in the velocity profile of the three-dimensional boundary layer which leads to cross-flow instability \citep{Reed1989, Saric2003}. Two types of cross-flow instability have been identified: traveling and stationary modes. The stationary mode plays an important role in the roughness-induced transition while traveling mode specifies external perturbations. The dominance of either type of modes depends on the specific configurations and the disturbance environment.

The global stability analyses (GSA) around a swept parabolic body in hypersonic flows were performed by \citet{Mack2008} and \citet{Mack2011a}. They were able to uncover a global spectrum containing boundary layer modes, acoustic modes and wave-packet modes. For the first time, these authors not only addressed the attachment-line instabilities, but also showed the connection of attachment line modes with cross-flow instability through GSA, though \citet{Bertolotti1999} had envisaged such connection in local stability analysis. Recently, \citet{Meneghello2015} performed global stability, receptivity and sensitivity analyses for incompressible flows around a Joukowski airfoil. They found that the global eigenfunction was the most responsive to forces applied in the vicinity of the attachment line. However, the analyses of such flows in hypersonic region are quite limited.

The receptivity process describes the procedure of penetration of external perturbations into the boundary layer and excitation of modes inside the boundary layer, which plays an important role in the boundary layer transition. Based on theoretical methods, such as finite-Reynolds-number methods \citep{Choudhari1994} and triple-deck theory \citep{Ruban1984}, the bi-orthogonal eigenfunction system was found to be an effective tool for the local receptivity analyses \citep{Hill1995, Fedorov2002, Tumin2007}. And a comprehensive review of this method is given by \citet{Tumin2020}. However, there exists no application of this approach to complex hypersonic flows. 

The objective of the present work is to globally extend the bi-orthogonal eigenfunction system to hypersonic sweep flow region, to highlight Mach-number effect and the receptivity features around the leading-edge of a swept blunt body. This paper is organized as follows. In section \S \ref{S2}, the governing equations are introduced and the bi-orthogonal eigenfunction system for global stability system is established for solving receptivity problem. The results for global stability/receptivity analysis are presented in section \S \ref{S3}. And the conclusion is given in section \S \ref{S4}.   

\begin{figure}
\begin{center}
\begin{tabular}{c}
\begin{overpic}[scale=0.35,tics=5]{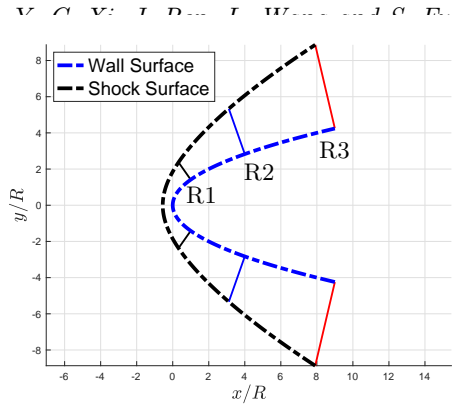}
\put(40,45){R1}
\put(53,50){R2}
\put(70,55){R3}
\end{overpic}
\end{tabular}
\caption{Outline of the computational domains of sweep parabolic body (\(x = y^2/2R)\). \(R\) represents the radius of the leading-edge. Three domains, bounded by the wall surface, shock surface and solid lines, which are marked by \(R1\), \(R2\) and \(R3\) from the smallest to the largest, are used for calculation of direct/adjoint eigenvalue problems. The largest region R3 is also used for basic flow calculations.}
\end{center}
\label{Figure1}
\end{figure}

\section{Theoretical framework}\label{S2}
The stability analyses of hypersonic steady flows is performed around the front part of a blunt body, as sketched in figure \ref{Figure1}. In general, the compressible Navier-Stokes equations for variable $\mathbf{Q} = (\rho,u,v,w,T)$, denoting the density, velocity components and temperature, can be written as 
\begin{equation}\label{eq1}
\bm{\Gamma}\frac{\partial \mathbf{Q}}{\partial t} = \mathscrbf{R}(\mathbf{Q}).
\end{equation}
The linearized Navier-Stokes equations (LNSE) around a stationary state $\mathbf{Q}$, with $\mathscrbf{R}(\mathbf{Q}) = 0$, can be represented by the combination of a linearized operator $\mathscrbf{L} = \partial{\mathscrbf{R}}/\partial\mathbf{Q}$ and a perturbation $\bm{p}=(\hat{\rho},\hat{u},\hat{v},\hat{w},\hat{T})$ field. This process forms a homogeneous system as 
\begin{equation}\label{eq2}
\bm{\Gamma}\frac{\partial \bm{p}}{\partial t}-\mathscrbf{L} \bm{p} = 0.
\end{equation} 
Taking a Laplace transform
\begin{equation}\label{eq3}
\hat{\bm{p}}(\omega) = \int_0^{\infty}\bm{p}(t)e^{-i\omega t}dt
\end{equation}
 in time leads to a standard global stability problem \citep{Theofilis2011}
\begin{equation} \label{eq4}
\left(-i \omega \bm{\Gamma} - \mathscrbf{L}\right)\hat{\bm{p}} = 0. 
\end{equation}
The detailed expressions of these operators are given in our previous work \citep{xi2020hypersonic}. For the fact that the operators are non-self-adjoint and the eigenfunctions are not orthogonal, a complete solutions of LNSE needs the help of adjoint eigenfunctions. An inner-product for two arbitrary vector \(\bm{a}\) and \(\bm{b}\) is defined as
\(
(\bm{a},\bm{b}) = \iint_{V}\bm{a}\cdot\bm{b}^{T}dxdy,
\)
and the superscript \(T\) stands for the transpose only.
Based on the inner-product, the adjoint problem of \eqref{eq4} can be expressed as:
\begin{equation} \label{eq5}
(-i\omega^{\dag} \bm{\Gamma}^{\dag} - \mathscrbf{L}^{\dag})\hat{\bm{p}}^{\dag} = 0,
\end{equation}
where the superscript \(\dag\) stands for the adjoint variables and operators.
It should be noted that boundary terms can be eliminated with proper boundary conditions during the process of integration by parts. 
In addition, based on the definition of inner-product, the adjoint problem and the original problem have the same eigenvalue spectrum. 
A bi-orthogonal eigenfunction system $\{ \hat{\bm{p}}_{a},\hat{\bm{p}}^{\dag}_{b}\}$ is formed with the help of eigenfunctions from the original problem \(\hat{\bm{p}}_a\) and the adjoint problem \(\hat{\bm{p}}^{\dag}_b\). By using the direct and adjoint equations, we can achieve the following relationship:
\begin{equation} \label{eq6}
 \left(\omega^{\dag}_b - \omega_a\right) \left( i\bm{\Gamma} \hat{\bm{p}}_{a},\hat{\bm{p}}^{\dag}_{b}  \right) = 0.                                                                                                 
\end{equation}
If \(\omega^{\dag}_b = \omega_a\) and further define that,
 \begin{equation} \label{eq7}
 \left(i\bm{\Gamma} \hat{\bm{p}}_{a},\hat{\bm{p}}^{\dag}_{b}\right) = C^{a}_{0},
 \end{equation}
then, \(C_0^{a}\) represents a normalization constant for a specific mode \(a\). If \(\omega^{\dag}_b \ne \omega_a\), the eigenfunctions are orthogonal to each other,
 \begin{equation} \label{eq8}
 \left(i\bm{\Gamma} \hat{\bm{p}}_{a},\hat{\bm{p}}^{\dag}_{b}\right) = 0.
 \end{equation}
Relations \eqref{eq7} and \eqref{eq8} form a bi-orthogonality condition in two/three dimensional domain. As most problems focus on the physical domain, an inverse Laplace transform
\begin{equation} \label{9}
\bm{p}(t) = \frac{1}{2\pi }\oint_{-\infty - i\infty}^{\infty + i\infty}\hat{\bm{p}}(\omega) e^{\omega t} d\omega,
\end{equation}
is adopted for the return to physical space.
Also, generally, the solution of linear Navier-Stokes equations can be expanded into the combination of continuous modes and discrete modes as
\begin{equation} \label{eq10}
\bm{p}(t) = \underbrace{ \sum_m \bm{C}_m^{d} \hat{\bm{p}}_m e^{-i\omega_m t} }_{Discrete \quad modes} + \underbrace{ \sum_n \oint \bm{C}_{n}^{c}(k)\hat{\bm{p}}_n(k) e^{-i\omega_n(k) t}dk }_{Continuous \quad modes},
\end{equation} 
where \(m\) and \(n\) represent the index of discrete modes and continuous branches, respectively; \(k\) stands for the integrate parameter along continuous branches, \(\bm{C}_m^{d}\) and \(\bm{C}_{n}^{c}\) represent the amplitude of those discrete modes and branches, respectively. 

With relationships \eqref{eq6}-\eqref{eq10}, for a specific discrete mode \(m\), we have
\begin{equation} \label{eq11}
\bm{C}_m^{d}e^{-i\omega_m t} = {\left(i\bm{\Gamma}\bm{p}_m,\hat{\bm{p}}^{\dag}_{m}\right)}/{C_0^{m}}.
\end{equation}
Thus, a generic external force \(\bm{f}\), representing small enough source term in the flow field, together with the external disturbances \(\bm{g}\) standing for any small surface/free-stream perturbations can be added to system \eqref{eq2}.
Then, the inhomogeneous system can now be expressed as
\begin{equation} \label{eq12}
\bm{\Gamma}\frac{\partial \bm{p}}{\partial t}-\mathscrbf{L} \bm{p} = \bm{f},
\end{equation}
with inhomogeneous boundary conditions \(\bm{g}\).
Similar to the previous process, a Laplace transform is adopted with respect to the force term and boundary conditions
\begin{equation} \label{eq13}
\hat{\bm{f}} = \int_{0}^{\infty}\bm{f}e^{-i\omega t}dt, \quad \hat{\bm{g}} = \int_{0}^{\infty}\bm{g}e^{-i\omega t}dt
\end{equation}
in time. The relative system in phase space can also be written as
\begin{equation} \label{eq14}
\left(-i \omega \bm{\Gamma} - \mathscrbf{L}\right)\hat{\bm{p}} = \hat{\bm{f}},
\end{equation}
with the inhomogeneous boundary conditions \(\hat{\bm{p}}=\hat{\bm{g}}\) at the boundary lines.

Considering a dot product between an adjoint eigenvector \(\hat{\bm{p}}^{\dag}\) and equation \eqref{eq14}, with integration over the whole domain, we can get the following identity for a specific discrete mode \(m\)
\begin{equation} \label{eq15}
i \left(\omega - \omega_m \right)\left( \bm{\Gamma} \hat{\bm{p}}_{m}, \hat{\bm{p}}^{\dag}\right) = \left(\hat{\bm{f}} , \hat{\bm{p}}^{\dag}\right) - B.C.
\end{equation} 
and
\begin{equation} \label{eq16}
\left( \bm{\Gamma}\bm{p}_{m},\hat{\bm{p}}^{\dag}\right) = \frac{1}{2\pi}\oint_{-\infty - i\infty}^{\infty + i\infty}\left(\bm{\Gamma}\hat{\bm{p}}_{m},\hat{\bm{p}}^{\dag}\right) e^{\omega t} d\omega
= \frac{1}{2\pi} \oint_{-\infty - i\infty}^{\infty + i\infty}\frac{\left(\hat{\bm{f}} , \hat{\bm{p}}^{\dag} \right) - B.C.}{i \left(\omega - \omega_m \right)}e^{\omega t}d\omega.
\end{equation}
\(B.C.\) term represents the concomitant boundary terms due to the inhomogeneous boundary conditions which are determined by using Green formulas and the detailed expression for this term is presented in appendix \ref{BCterms}.
By closing the Bromwich integral in the complex \(\omega\)-plane with residues theorem, the amplitude \(\bm{C}_m^{d}\) for a specific discrete mode \(m\) can be recoved as
\begin{equation} \label{eq17}
\bm{C}_m^{d} = \left|\left[{\left(\hat{\bm{f}} , \hat{\bm{p}}^{\dag}_{m} \right) - B.C. }\right]/{C^{m}_0}\right|. 
\end{equation}
For the fact that \(C_0^m\) is a constant for specific discrete mode, the adjoint field represents a scaled Green's function for receptivity problem. In fact, formula \eqref{eq17} is consistent with those obtained from incompressible flows \citep{Giannetti2007}. It can thus be used as a general form for incompressible or compressible flow system. 

\section{Global stability and receptivity of the leading-edge region} \label{S3}
The test model is a two-dimensional parabolic body swept with a sweep angle \(\Lambda\). The model geometry and computational domain are shown in figure \ref{Figure1}. The freestream Reynolds number \(Re_{\infty}\), the sweep Reynolds number \(Re_s\), the freestream Mach number \(M_{\infty}\), the sweep Mach number \(M_s\) and the recovery temperature \(T_r\) are defined as
\begin{equation}
\begin{aligned}
Re_{\infty} = \frac{|\vec{V}_{\infty}| \delta^*}{\nu_{\infty}},~Re_s = \frac{W_\infty \delta^*}{\nu_r},~M_{\infty} = \frac{|\vec{V}_{\infty}| }{c_{\infty}}=8.15, ~M_s = \frac{W_{\infty}}{c_s}, \\
 T_r = T_{\infty} + \sigma (T_0 - T_{\infty}), \textrm{where}~\sigma = 1 - (1 - \xi_w) \textrm{sin}^2\Lambda.
 \end{aligned}
\end{equation}
Here, $\xi_w$ is a constant for specific freestream conditions (\(M_{\infty}\) and \(\Lambda\)) and determined based on the study of \citet{Reshotko1958}; \(R=0.1m\) represents the radius of the leading-edge; \(\vec{V}_{\infty}\) stands for the freestream velocity vectors with \(U_{\infty},V_{\infty}\) and \(W_{\infty}\) along \(x,y\) and \(z\) direction, respectively. 
\(T_{\infty}\) of \(50.93K\) and \(T_{0}\) stand for the freestream and stagnation temperature, respectively.
The parameters $c_{\infty}$ and $c_s$ are the sound speed before and after the leading shock, \(\nu_r\) represents the kinematic viscosity at $T_r$. The viscosity length scale $\delta^*$ is defined as \(\delta^* = \sqrt{{\nu_r R}/{2 U_2}},\)
where \(U_2\) represents chord-wise velocity behind the shock.
The Prandtl number $Pr$ of \(0.71\) and the specific heat ratio \(\gamma\) of \(1.4\) are set following the ideal gas assumption of air. 
Table \ref{table1} lists the parameters of the chosen cases. 
Here, \(M_s\) is used to define the supersonic and hypersonic configurations as mentioned below. 
\(T_w = T_r\) is the surface temperature and is used for all cases. And the spanwise wave number \(\beta\) correspond to the physical wavelength \(\lambda^*_z\) of \(11.7\)mm are used.

\begin{table}
\begin{center}
\begin{tabular}{ccccccc}
                & $\Lambda(^o)$ & $\delta^*(m)$ & $Re_{\infty}$ & $Re_{s} $ & $M_s$ & $\beta$\\  
P1 &35.0 & 2.0150e-4 & 2826.27 &  937.33 & 1.5081 &  0.1079 \\
P2 &40.0 & 2.0746e-4 & 2909.82 &1075.44 & 1.7948 &  0.1111 \\
P3 &45.0 & 2.1492e-4 & 3014.45 &1213.22 & 2.1188 &  0.1151 \\
P4 &50.0 & 2.2429e-4 & 3145.92 &1348.36 & 2.4922 &  0.1201 \\
P5 &55.0 & 2.3617e-4 & 3312.54 &1476.33 & 2.9313 &  0.1265 \\
P6 &60.0 & 2.5145e-4 & 3526.78 &1588.28 & 3.4585 &  0.1347 \\
P7 &65.0 & 2.7151e-4 & 3808.15 &1666.99 & 4.1039 &  0.1454 \\
P8 &70.0 & 2.9868e-4 & 4189.29 &1679.10 & 4.9055 &  0.1600                 
\end{tabular}
\end{center}
\caption{Flow parameters of eight cases in the present study.}
\label{table1}
\end{table}%

Firstly, a high-order shock fitting method is used to obtain the basic flow fields. A matrix-based high-order global stability analysis is then performed to solve the direct and adjoint problems. In the shock fitting method, the shock is modeled as a boundary of the computational domain, smooth fields can be achieved for the adjoint calculation without considering shock discontinuities \citep{Giles2001}. Details of the numerical approaches and solver validation/verification can be found in \citet{xi2020hypersonic}.  To achieve a grid-independent solution, a large number of grid points up to \(5401\times401\) are employed to identify the structure of direct and adjoint eigenfunctions with a minimum of 20 points per wavelength in each direction for different sizes of computational domains.  

\begin{figure}
\begin{center}
\begin{tabular}{ccc}
\begin{overpic}[scale=0.319,tics=5]{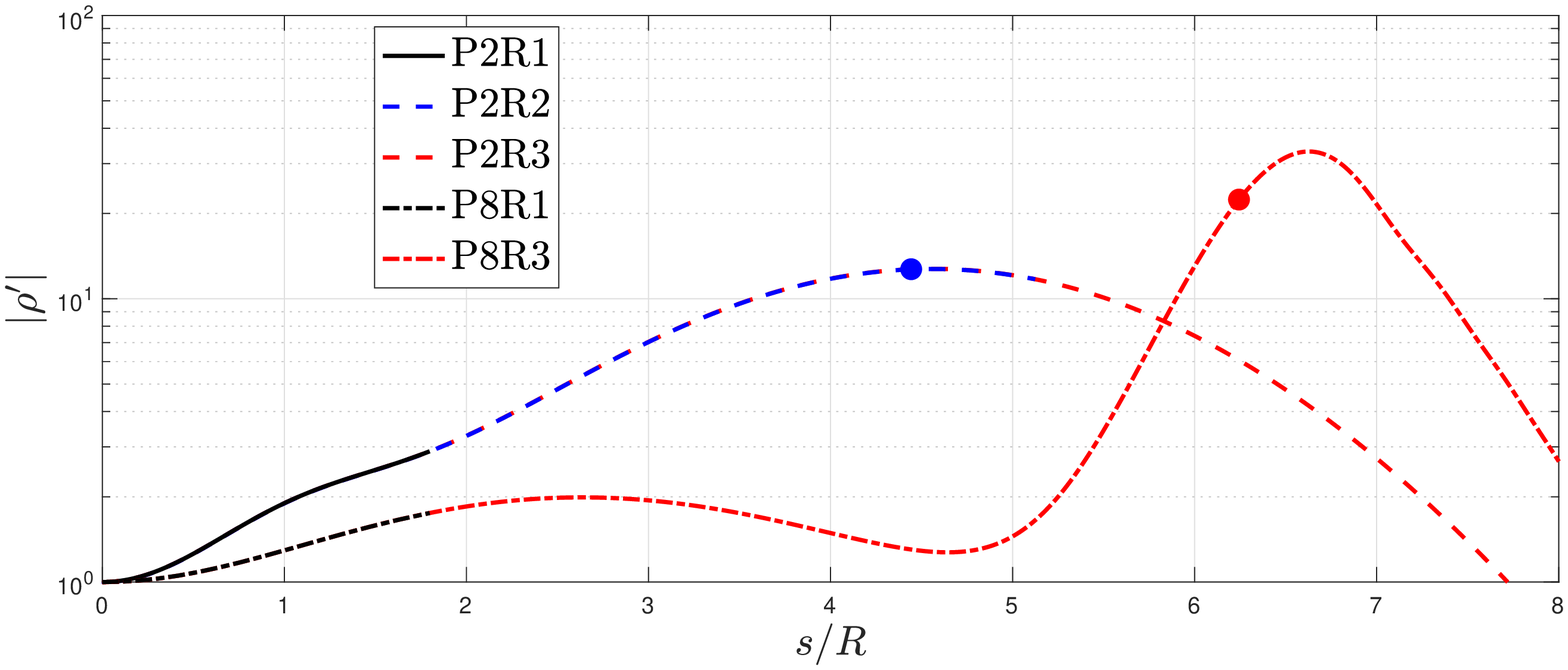}
\put(10,35){$(a)$}
\put(55,30){$m1$}
\put(75,34){$m2$}
\end{overpic}
\begin{overpic}[scale=0.343,tics=5]{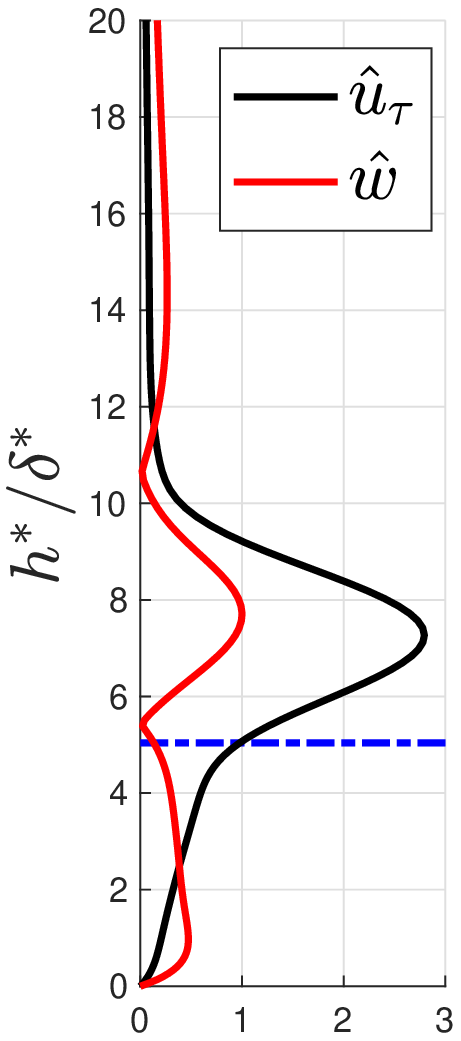}
\put(-3,90){$(b)$}
\end{overpic}
\begin{overpic}[scale=0.343,tics=5]{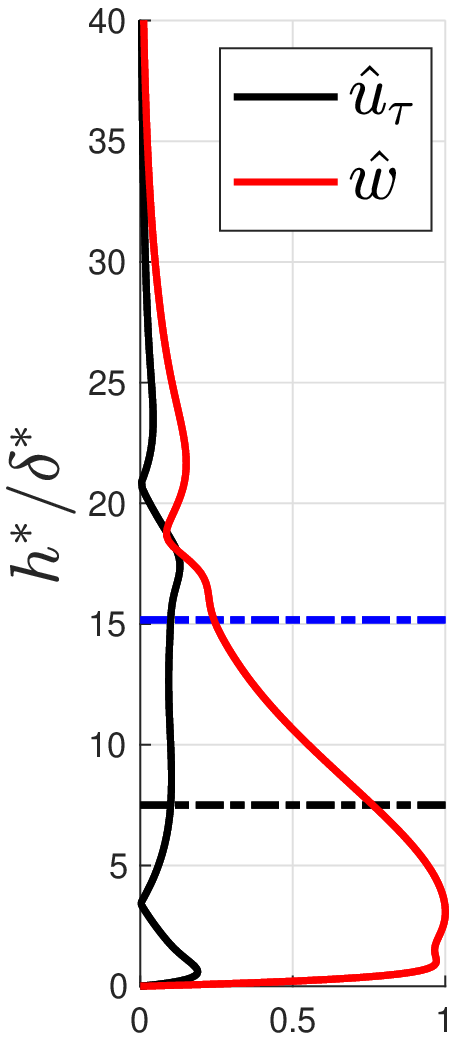}
\put(-2,90){$(c)$}
\end{overpic}
\end{tabular}
\end{center}
\begin{center}
\begin{tabular}{ccc}
\hline
  Small domain (P2R1)  &  Mid-sized domain (P2R2) & Large domain (P2R3) \\
$0.035724347+0.00011250136i$ & 
$0.035724347+0.000112501\underline{49}i$ &
$0.035724347+0.000112501\underline{56}i$ \\
  Small domain (P8R1)  &  Mid-sized domain (P8R2)  & Large domain (P8R3) \\
$0.12482622 + 0.00015009545i$& $0.1248262\underline{3} + 0.00015009\underline{213}i$ &$0.12482622 + 0.00015009\underline{117}i$ \\
\hline
\end{tabular}                   
\end{center}
\caption{$(a)$. Dependences of the density perturbations along wall surfaces on different computational domain size (R1, R2 and R3 shown in figure \ref{Figure1}) for CASES P2 and P8. All the perturbations are normalized with the values at the attachment line. Computed direct eigenvalues for CASES P2 and P8 over different domains are shown in the table. The digits different from the small-domain values are marked underline. The two eigenfunctions at \(m1\) and \(m2\) in $(a)$ are shown in \((b)\) and \((c)\). \(\hat{u}_{\tau}\) and \(h^*\) stand for chordwise velocity perturbation and dimensional distance away from wall surface. And \(s\) represent the distance away from the attachment line along the surface. The location of critical layer \(c_r = \overline{U}\) and the sonic lines \(c_r = \overline{U}+a\) are indicated by horizontal dash-dotted blue and black lines, respectively.}
\label{Figure2}
\end{figure}

\begin{figure}
\begin{center}
\begin{tabular}{cc}
\begin{overpic}[scale=0.29,tics=5]{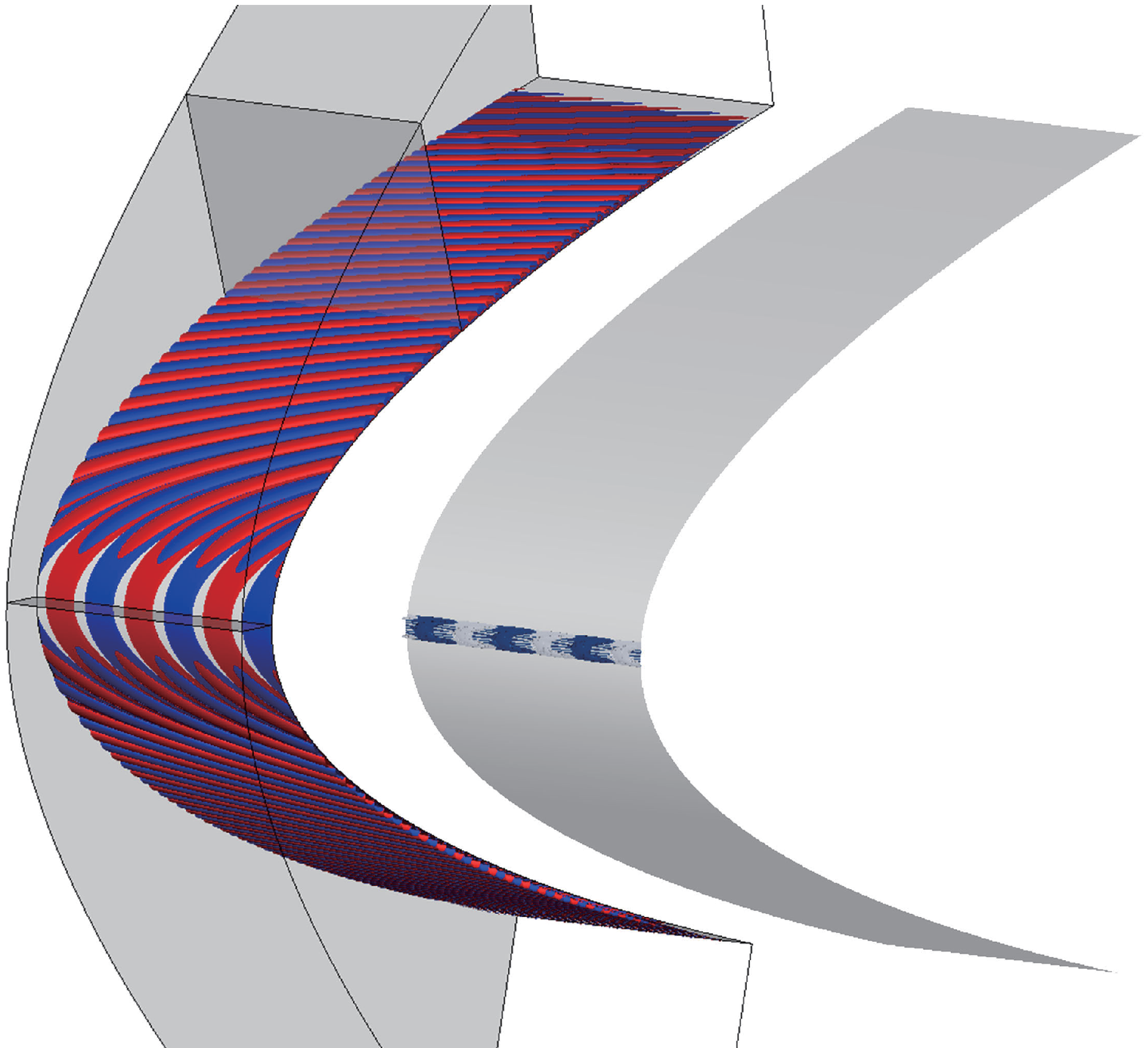}
\put(5,78){$(a)$}
\put(30,35){$a1$}
\put(45,55){$a2$}
\put(5,10){Direct}
\put(81,10){Adjoint}
\put(81,30){P2}
\end{overpic} &
\begin{overpic}[scale=0.31,tics=5]{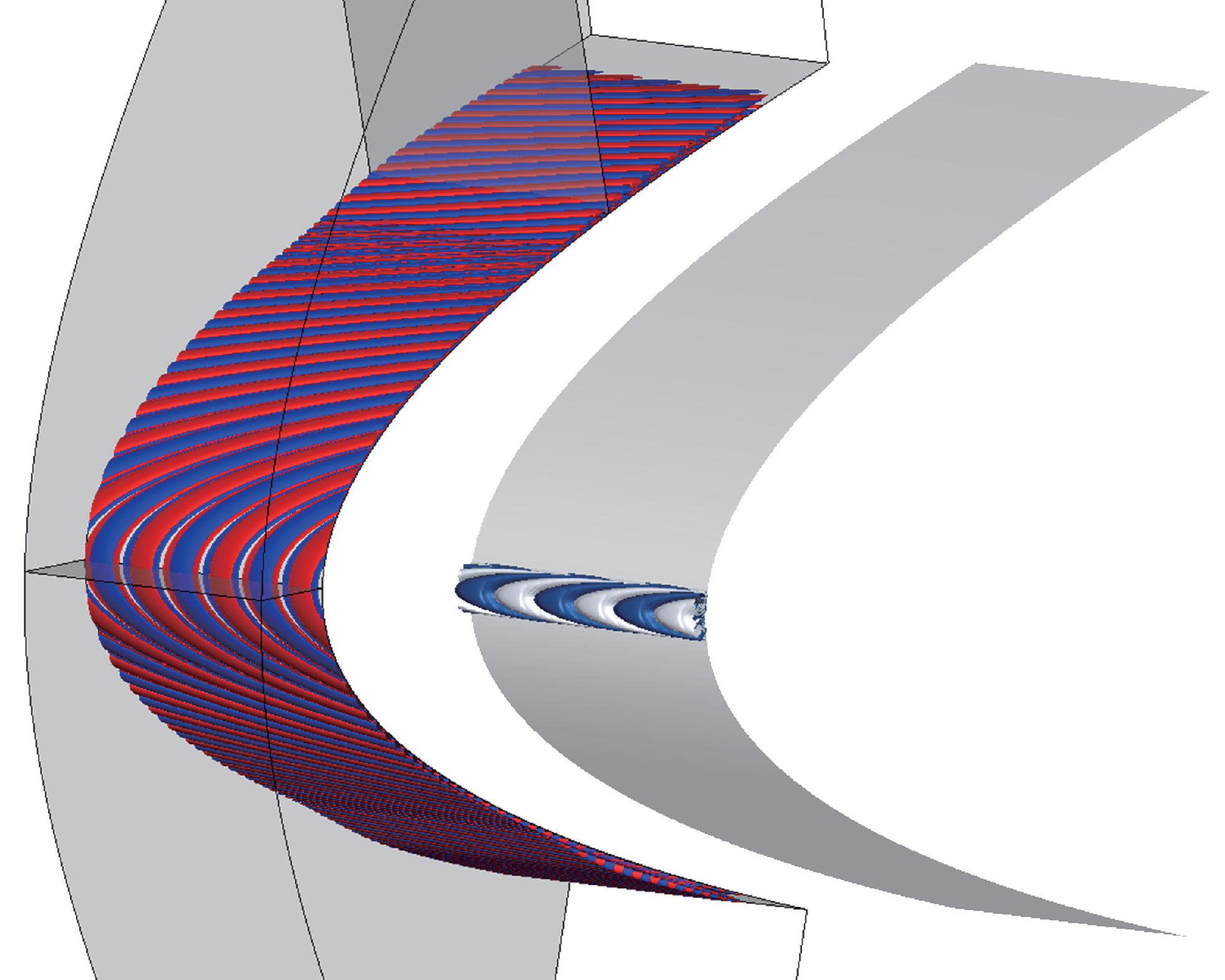}
\put(1,73){$(b)$}
\put(30,35){$b1$}
\put(55,60){$b2$}
\put(5,10){Direct}
\put(80,10){Adjoint}
\put(81,30){P8}
\end{overpic} \\
\begin{overpic}[scale=0.31,tics=5]{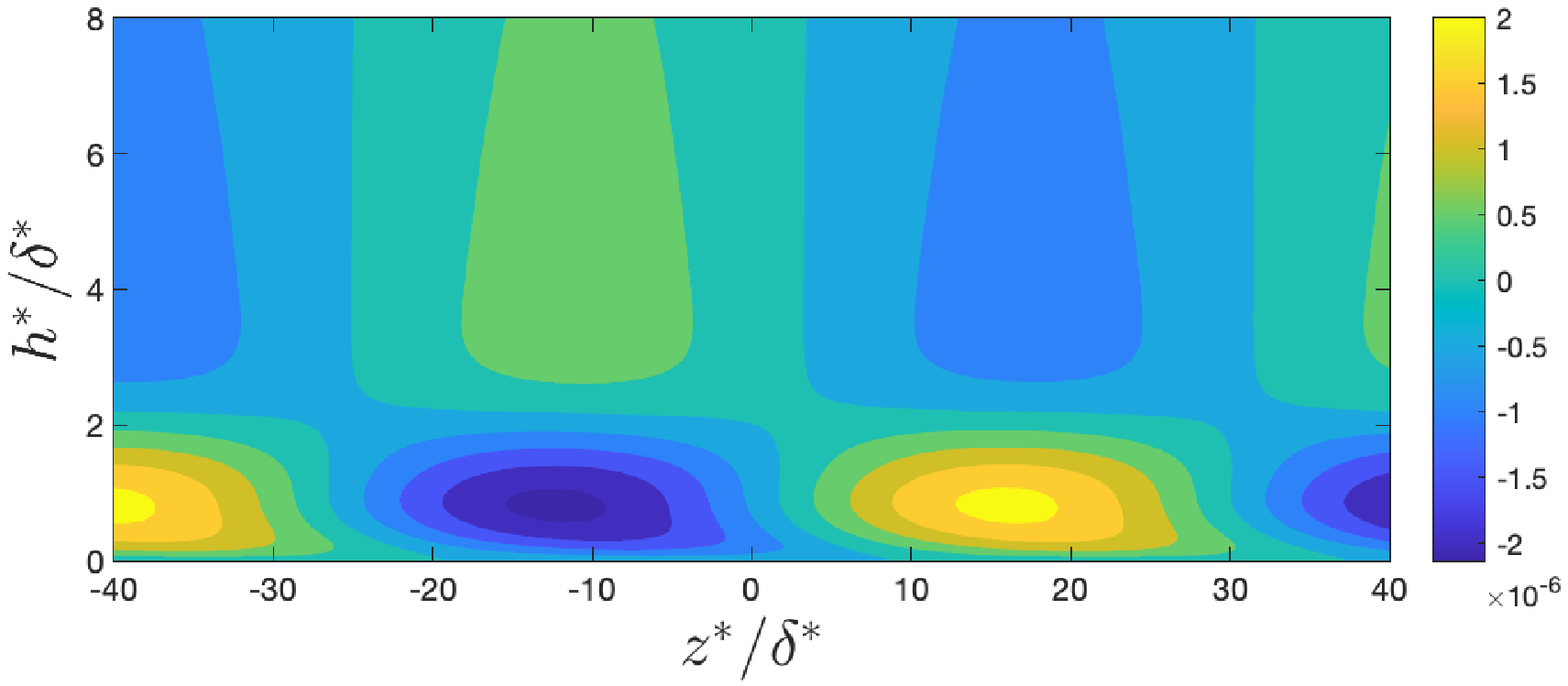}
\put(1,33){$(c)$}
\put(94,35){$\hat{w}$}
\end{overpic} &
\begin{overpic}[scale=0.31,tics=5]{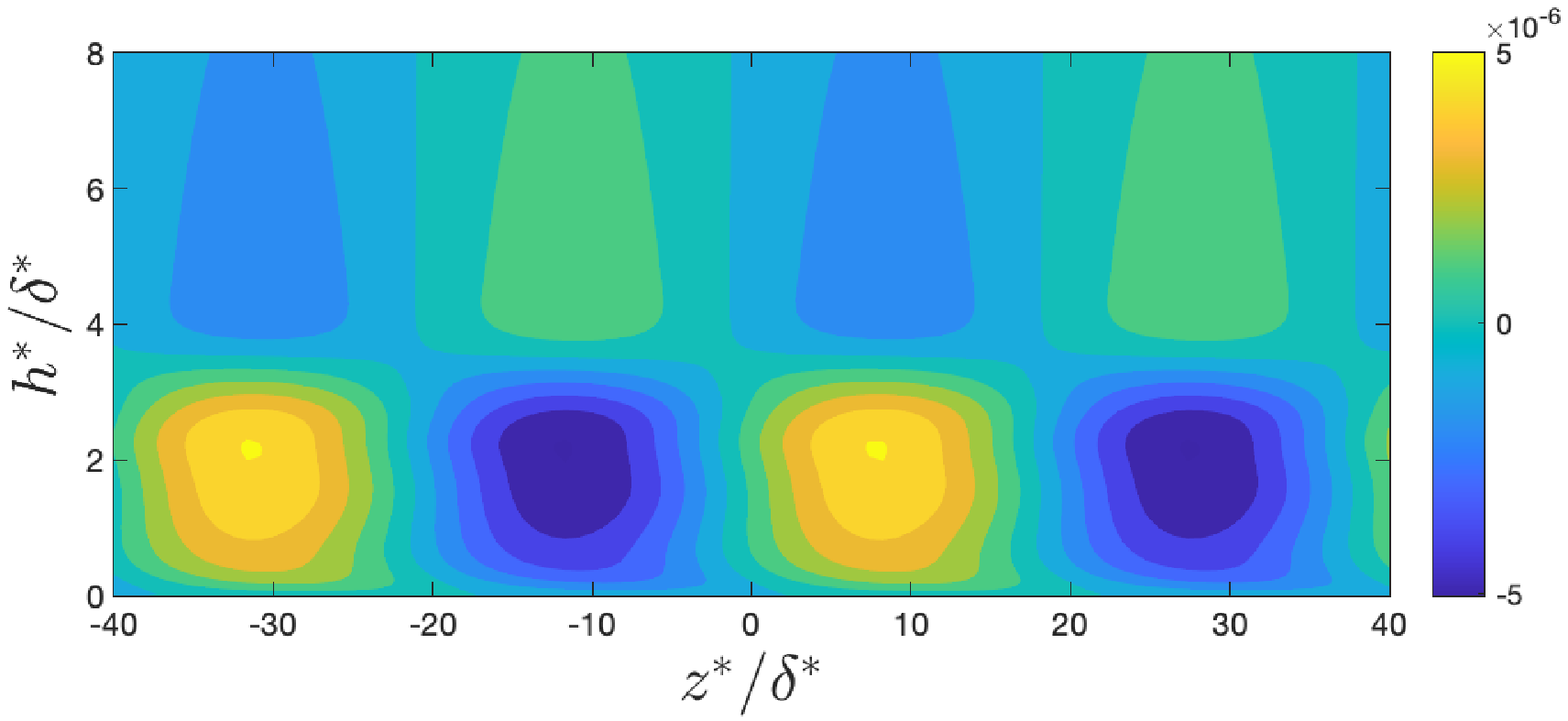}
\put(1,33){$(d)$}
\put(94,35){$\hat{w}$}
\end{overpic} \\
\begin{overpic}[scale=0.31,tics=5]{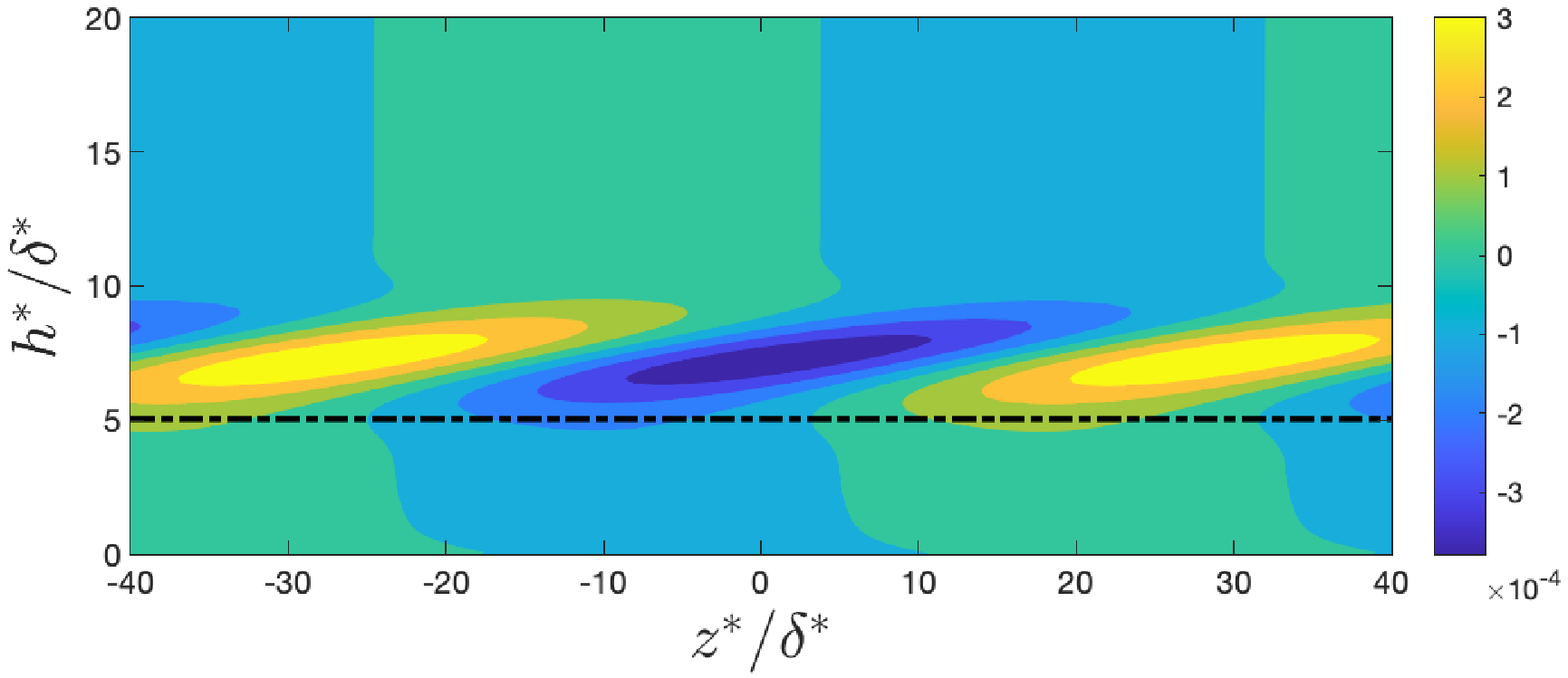}
\put(1,33){$(e)$}
\put(50,15){Critial layer}
\put(94,35){$\hat{u}_{\tau}$}
\end{overpic} &
\begin{overpic}[scale=0.31,tics=5]{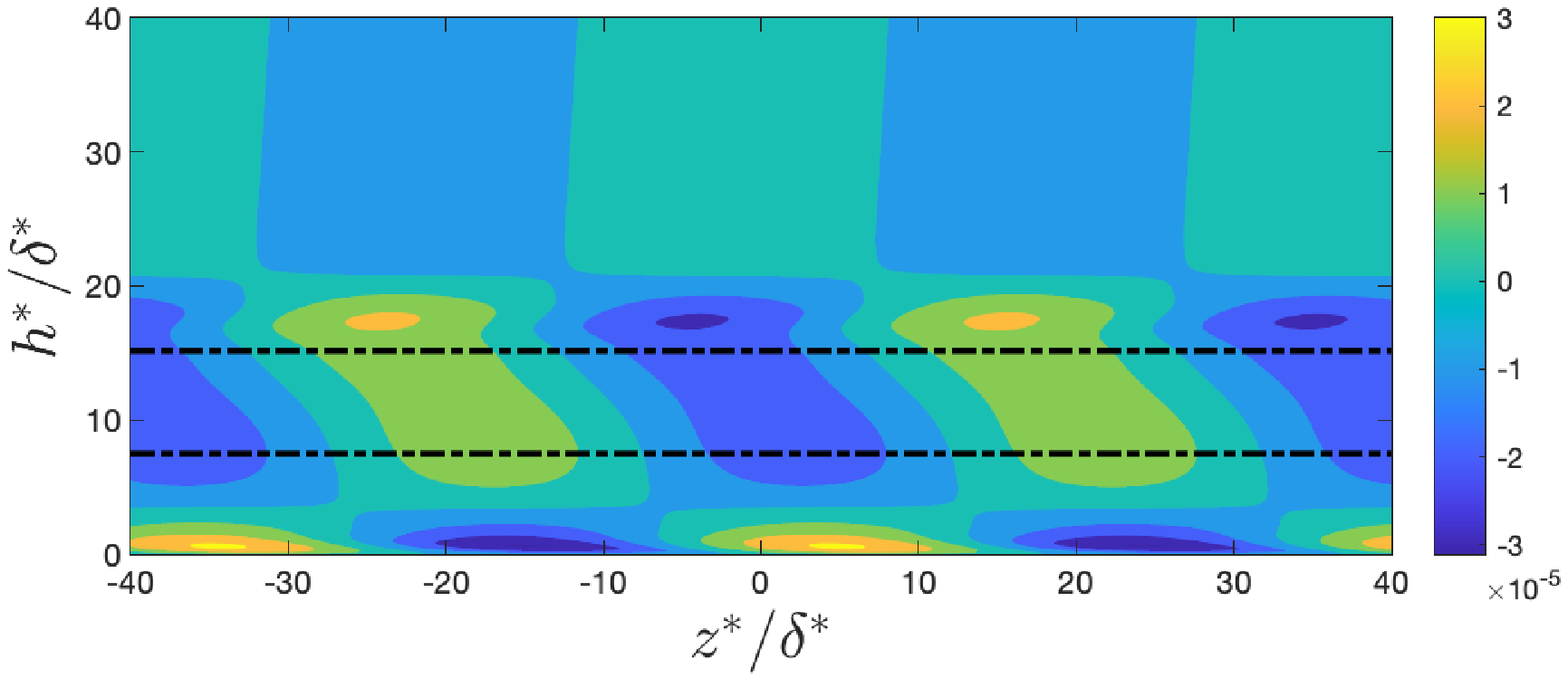}
\put(1,33){$(f)$}
\put(50,20){Critial layer}
\put(50,14){Sonic line}
\put(94,35){$\hat{u}_{\tau}$}
\end{overpic} \\
\begin{overpic}[scale=0.31,tics=5]{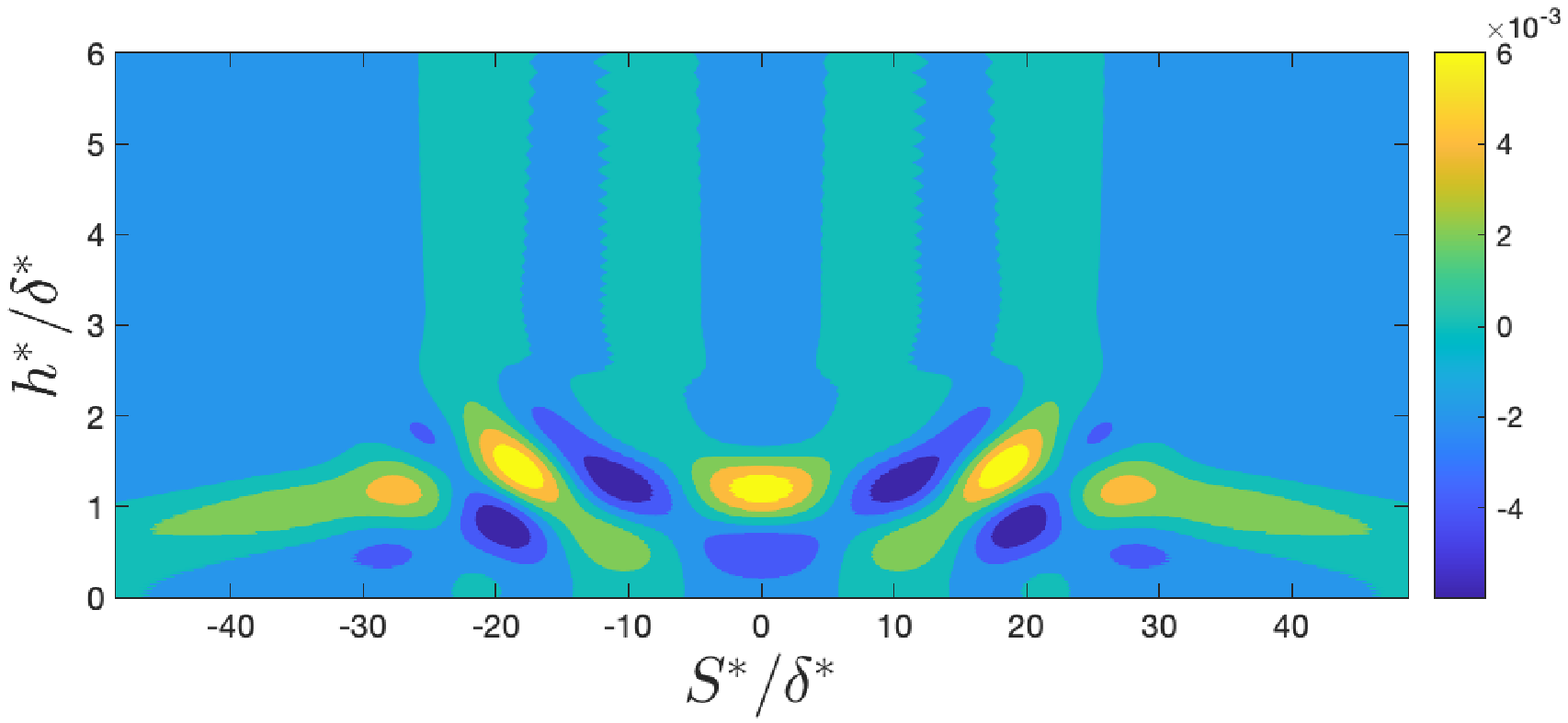}
\put(1,33){$(g)$}
\put(94,35){$\hat{w}^{\dag}$}
\end{overpic} &
\begin{overpic}[scale=0.31,tics=5]{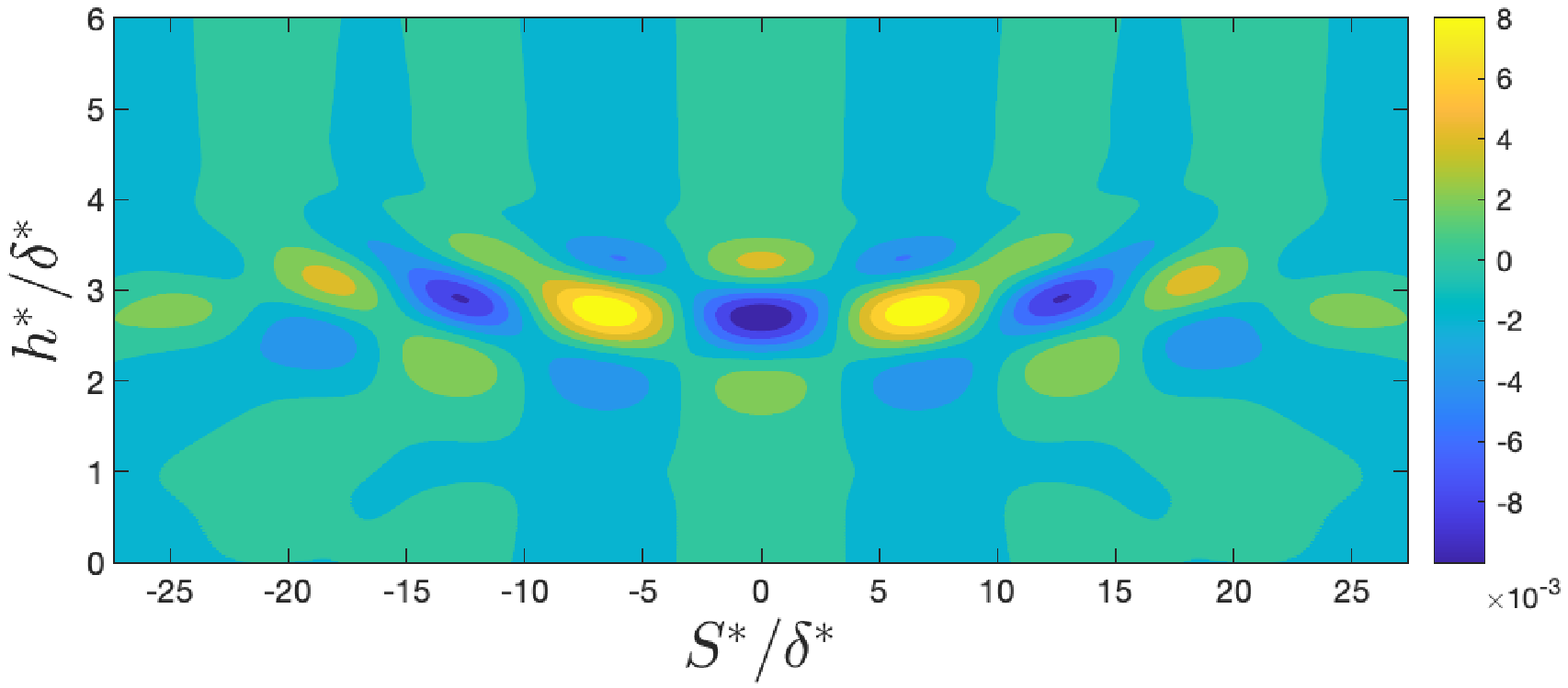}
\put(1,33){$(h)$}
\put(94,35){$\hat{w}^{\dag}$}
\end{overpic}
\end{tabular}
\end{center}
\caption{$(a)$ The direct eigenfunction of the mode for CASE P2 is visualized on the left by iso-surfaces (red/blue for positive/negative values for direct eigenfunctions, $\pm10^{-6}$) of the real part of the spanwise velocity \(\hat{w}\). Two slices $a1$ (attachment-line plane) and $a2$ (further downstream plane) are marked and shown in $(c)$ with \(\hat{w}\) at the attachment line and $(e)$ with the chordwise velocity \(\hat{u}_{\tau}\) in the cross-flow region, respectively. The adjoint eigenfunctions of this mode is also shown in \((a)\), on the right, with the iso-surface (blue/white for positive/negative values, $\pm 10^{-3}$) of the adjoint spanwise velocity \(\hat{w}^{\dag}\) and the leading edge region of adjoint field along the chord-wise direction is shown in \((g)\). $(b)$ The direct eigenfunction of the mode for CASE P8 is visualized on the left by iso-surfaces of the real part of the spanwise velocity \(\hat{w}\). Two slices $b1$ and $b2$ are marked and shown in $(d)$ with the spanwise velocity and $(f)$ with the chordwise velocity, respectively. The adjoint eigenfunctions of this mode is also shown in \((b)\), on the right, with the iso-surface of the adjoint spanwise velocity and the leading edge region of adjoint field along the chord-wise direction is shown in \((h)\). \(h^*\) and \(z^{*}\) represent the dimensional distance away from wall surfaces and spanwise location. \(S^*\) represents the distance away from the attachment line along the surface, and the positive/negative values distinguish the upper\((y>0)\) and lower part\((y<0)\) of the field.}
\label{Figure3}
\end{figure}

The leading attachment-line mode is calculated through three different domains (see figure \ref{Figure1}). For all domains, we obtain the same leading eigenvalues and eigenfunctions as presented in figure \ref{Figure2}. It indicates that the characteristics of attachment-line modes are not affected by the computation domain size. From figure \ref{Figure2}$(a)$, the behavior of the surface density perturbations are of great difference in the downstream region \(s/R>1\) for these two cases. Based on eigenfunctions in figure \ref{Figure2}\((b)\) and \((c)\), it is shown that the downstream perturbations in CASE P2 are of the cross-flow type while those in CASE P8 are of the second Mack-mode type, which can be seen more clearly in Figure \ref{Figure3}$(e)$ and $(f)$.

Figure \ref{Figure3} displays the direct and adjoint eigenvectors, obtained from the largest R3 domain, for the leading eigenvalues of CASES P2 and P8 with the iso-surfaces of the real part for the spanwise velocity \(\hat{w}\). Figure \ref{Figure3}$(a)$ and $(b)$ shows the typical features for the direct global eigenfunctions which exhibits both attachment-line and cross-flow instability (CASE P2)/the second Mack mode instability (CASE P8). For CASE P2 with small \(M_s\), from the attachment-line plane (figure \ref{Figure3}$(c)$) to the further downstream plane (figure \ref{Figure3}$(e)$), perturbations evolve away from wall surfaces and then form cross-flow vortices aligning with the external streamlines, as also reported by \citet{Mack2008} and \citet{Meneghello2015}. While for CASE P8 with large \(M_s\), despite the similar mode structure at the attachment-line plane (figure \ref{Figure3}$(d)$), the perturbations at the downstream plane (figure \ref{Figure3}$(f)$) exhibit the feature of the second Mack mode instability in which the perturbations are mainly located below the sonic line. In contrast to the feature that the perturbations of direct mode within large area, the adjoint mode only appears in the vicinity of the attachment line, as shown in figure \ref{Figure3}$(a)$, \((b)\), $(g)$ and $(h)$. Moreover, based on our theoretical receptivity analysis on external force \(\bm{f}\), this region is the most responsive to this discrete global mode. In other words, even the direct global mode covers a large region further away from the attachment line we can still control/excite the mode by introducing forces at a relatively small region close to the attachment line. This finding consists with those from incompressible flows by \citet{Meneghello2015}, and also indicates that the leading edge control may have a great effect on the sweep hypersonic blunt bodies.

\begin{figure}
\begin{center}
\begin{tabular}{cc}
\begin{overpic}[scale=0.31,tics=5]{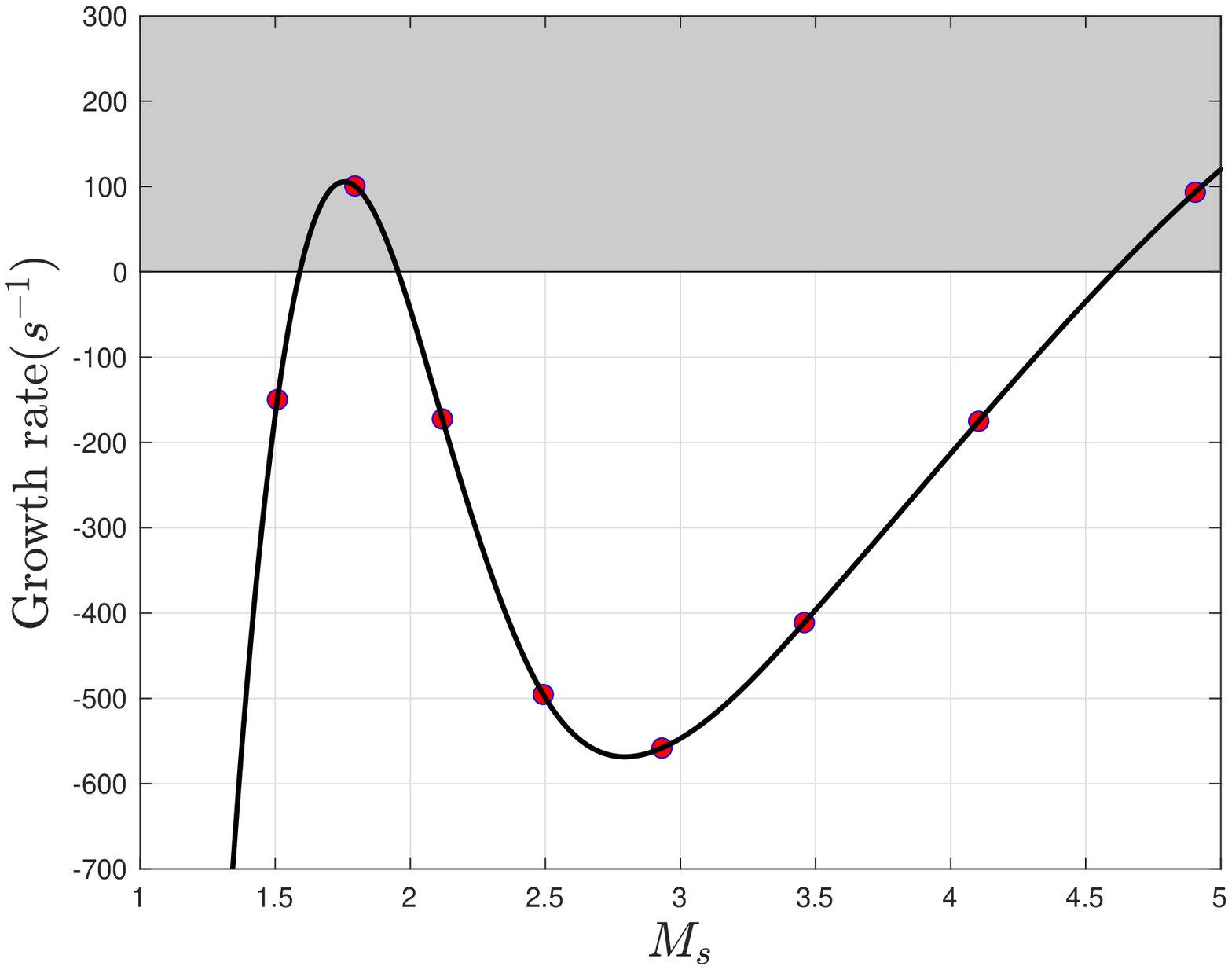}
\put(30,70){$(a)$}
\put(50,70){Unstable}
\put(55,40){Stable}
\end{overpic} &
\begin{overpic}[scale=0.31,tics=5]{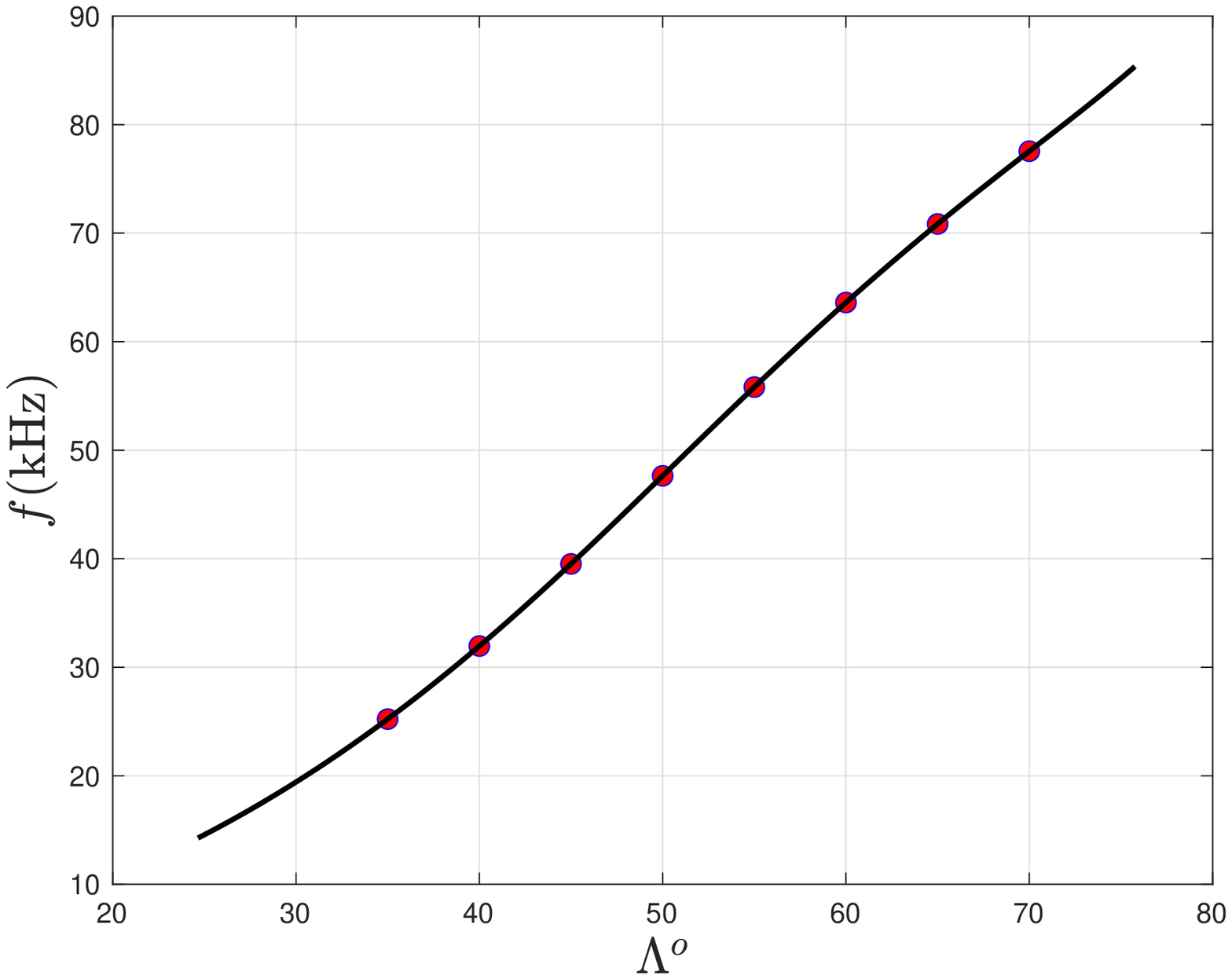}
\put(30,70){$(b)$}
\end{overpic}
\end{tabular}
\end{center}
\caption{\((a)\) Dependence of the growth rates on sweep Mach numbers. \((b)\) Dependence of the frequency on sweep angle \(\Lambda\). The marked red points represent CASES P1-P8.}
\label{Figure4}
\end{figure}

The characteristics of the leading attachment-line mode on sweep angles and the relative sweep Mach numbers is shown in figure \ref{Figure4}. As the sweep angle varies from \(20\) to \(70\) degrees, the growth rate of the leading attachment-line mode shows a trend from rising to decline, reaches the locally highest point around \(40\) degree. When the sweep angle further increases, the growth rate of the leading attachment-line mode exhibits the features of monotonous increase and becomes unstable at large sweep Mach numbers. Moreover, this finding is consistent with two facts: the first one is that the plane stagnation flow is known to be linearly stable to three-dimensional perturbations, which can be seen as the limit of zero sweep angle for the present cases; the other is that the attachment-line instability is found to become unstable for large sweep angles \citep{Gaillard1999}.

\begin{figure}
\begin{center}
\begin{overpic}[scale=0.25,tics=5]{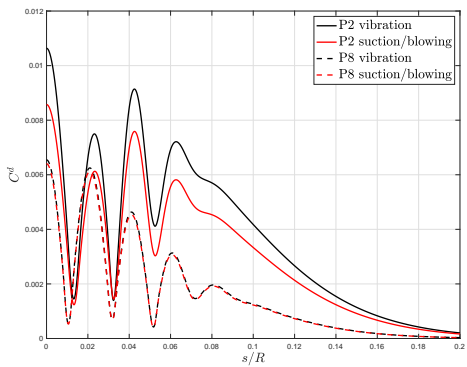}
\end{overpic}
\end{center}
\caption{Amplitudes of the leading attachment-line modes excited at different location \(s/R\) for surface vibration and suction/blowing of delta type. All the eigenfunctions are normalized so that the density components on the wall surface are equal to one at the attachment line.}
\label{Figure5}
\end{figure}

Based on formula \eqref{eq17}, the amplitudes of the modes excited by various boundary perturbations are mainly determined by the concomitant boundary terms (\(B.C.\)). From the expression in appendix \ref{BCterms}, it is clearly shown that the receptivity is evaluated from the gradients for adjoint variables and physical perturbations on the boundary. Moreover, this procedure permits the extraction of the receptivity amplitude pertaining to any discrete modes for any type of boundary perturbations at any location. Taking the surface perturbations in CASES P2 and P8 for instances, two types of surface perturbation (a delta function form perturbation with subcritical frequency) are calculated, as shown in figure \ref{Figure5}. It demonstrates that the strong receptive regions of the leading attachment-line modes to surface perturbations are in the vicinity of attachment line and the surface perturbations at the exact attachment line are the most effective. Also, the surface vibrations are found to be a little more effective than suction/blowing, because of the higher excited amplitudes. Moreover, as the delta-function form perturbations are used and the basis for representation of LNSE cover the whole domain, the distributed surface perturbations receptivity analyses can be easily performed by integrations along the finite length. Further, as the present framework is from the basis of LNSE, the receptivity problem for the excitation of discrete modes of freestream disturbances with surface inhomogeneities can be also taken into account.

As mentioned by \citet{Tumin2020}, the eigenfunction expansion methods for the two-dimensional and three-dimensional problems are the natural extensions of the widely used local cases. The analysis of discrete modes is similar for both local and global cases and can be easily performed. However, things become much more complex with continuous spectrum, as the feature of a linear ordinary operator and partial operator is quite different. Although there exist uncertainties of the continuous spectra for global cases, the bi-orthogonal eigenfunction system is still a powerful tool for solving global receptivity problems to discrete modes for complex high-speed boundary layers.

\section{Conclusions}\label{S4}
Global instabilities and receptivities around leading-edge of a swept blunt body are studied in hypersonic flow region. The characteristics of the leading attachment-line mode to the variation of sweep angles from 20 to 70 degrees are obtained for the first time. As the sweep angle increases, the growth rates of the leading attachment-line modes exhibit a trend from rising to decline and rising again in which the modes also show the transformation for the features of the cross-flow instability to the second Mack mode instability further downstream.
Moreover, a general bi-orthogonal eigenfunction system for global stability system is established to address receptivity problems to any external forces and boundary perturbations. The receptivity analyses indicate that the leading attachment-line mode is the most responsive to external disturbances in the vicinity of leading edge. It also clarifies that the framework is simple and can be easily applied for local and non-local linear stability/receptivity analyses with real geometries.  

\begin{acknowledgments}
We appreciate Dr. Meelan Choudhari and Prof. Ardeshir Hanifi for very useful discussion on receptivity analysis. This work received partial support from National Key Project GJXM92579, National Sci.\ \& Tech.\ Major Project (2017-II-0004-0016).
\end{acknowledgments}

\section{Declaration of Interests}
 The authors report no conflict of interest.
 
\appendix
\section{Concomitant boundary terms}\label{BCterms}
The concomitant boundary terms are determined by using Green formulas. Details for derivation of these terms can be found in supplement material. Taking the detailed explicit forms of the operators \(\mathscrbf{L}\), the \(B.C.\) term can be expressed as:
\begin{align}
B.C. &= -\int_{\Gamma}\rho^{\dag}\rho \hat{u}_{bc} dy + \int_{\Gamma}\rho^{\dag}\rho \hat{v}_{bc} dx  \\ \nonumber
&- \int_{\Gamma} \left(\frac{4\mu}{3Re}\frac{\partial u^{\dag}}{\partial x}\hat{u}_{bc} +\frac{\mu}{Re}\frac{\partial v^{\dag}}{\partial x}\hat{v}_{bc} + \frac{\mu}{Re}\frac{\partial w^{\dag}}{\partial x}\hat{w}_{bc} + \frac{\mu}{RePr}\frac{\partial T^{\dag}}{\partial x}\hat{T}_{bc}\right)dy \\ \nonumber
&+\int_{\Gamma} \left(\frac{\mu}{Re}\frac{\partial u^{\dag}}{\partial y}\hat{u}_{bc} +\frac{4\mu}{3Re}\frac{\partial v^{\dag}}{\partial y}\hat{v}_{bc} + \frac{\mu}{Re}\frac{\partial w^{\dag}}{\partial y}\hat{w}_{bc} + \frac{\mu}{RePr}\frac{\partial T^{\dag}}{\partial y}\hat{T}_{bc}\right)dx \\ \nonumber
&- \frac{1}{2}\int_{\Gamma} \left[\frac{\mu}{3Re}\frac{\partial v^{\dag}}{\partial y}\hat{u}_{bc} + \frac{\mu}{3Re}\frac{\partial u^{\dag}}{\partial y}\hat{v}_{bc}\right]dy
+ \frac{1}{2}\int_{\Gamma} \left[\frac{\mu}{3Re}\frac{\partial v^{\dag}}{\partial x}\hat{u}_{bc} + \frac{\mu}{3Re}\frac{\partial u^{\dag}}{\partial x}\hat{v}_{bc}\right]dx,
\end{align}
where the superscript \(\dag\) and subscript \(bc\) represent the adjoint variables and variables at the boundary line \(\Gamma\). The direction of the boundary line \(\Gamma\) is defined to ensure the computation domain is always on the left. The linearized boundary conditions over the surface for small enough surface vibrations can be specified as
\begin{subequations}
\begin{equation}
\left(\hat{u}_{bc}, \hat{v}_{bc}, \hat{w}_{bc}, \hat{T}_{bc}\right) ^{T} =
H(x,y)\left(\hat{u}_{bc}^{\prime}, \hat{v}_{bc}^{\prime}, \hat{w}_{bc}^{\prime}, \hat{T}_{bc}^{\prime}\right)^{T}  
\end{equation}
\begin{equation}
\hat{u}^{\prime}_{bc} = -\frac{\partial u}{\partial \bm{n}} + \underline{\text{i}\omega \bm{n}_x},\quad
\hat{v}^{\prime}_{bc} = -\frac{\partial v}{\partial \bm{n}} + \underline{\text{i}\omega \bm{n}_y},\quad
\hat{w}^{\prime}_{bc} = -\frac{\partial w}{\partial \bm{n}},\quad
\hat{T}^{\prime}_{bc} = -\frac{\partial T}{\partial \bm{n}}, 
\end{equation}
\end{subequations}
where \(H(x,y)\) is the shape function of disturbances. And \(\bm{n}\) stands for surface normal direction and the parts underlined represent the only terms for blowing/suction.

\bibliographystyle{jfm}
\bibliography{Reference}

\end{document}